# Impact of PSF misestimation and galaxy population bias on precision shear measurement using a CNN

L. M. Voigt ⋆

*School of Mathematics, Statistics and Actuarial Science, University of Essex, Wivenhoe Park, Colchester, CO4 3SQ*

**ABSTRACT**
Weak gravitational lensing of distant galaxies provides a powerful probe of dark energy. The aim of this study is to investigate the application of convolutional neural networks (CNNs) to precision shear estimation. In particular, using a shallow CNN, we explore the impact of point spread function (PSF) misestimation and 'galaxy population bias' (including 'distribution bias' and 'morphology bias'), focusing on the accuracy requirements of next generation surveys. We simulate a population of noisy disc and elliptical galaxies and adopt a PSF that is representative of a *Euclid*-like survey. We quantify the accuracy achieved by the CNN, assuming a linear relationship between the estimated and true shears and measure the multiplicative ($m$) and additive ($c$) biases. We make use of an unconventional loss function to mitigate the effects of noise bias and measure $m$ and $c$ when we use either: (i) an incorrect galaxy ellipticity distribution or size–magnitude relation, or the wrong ratio of morphological types, to describe the population of galaxies (distribution bias); (ii) an incorrect galaxy light profile (morphology bias); or (iii) a PSF with size or ellipticity offset from its true value (PSF misestimation). We compare our results to the *Euclid* requirements on the knowledge of the PSF model shape and size. Finally, we outline further work to build on the promising potential of CNNs in precision shear estimation.

**Key words:** gravitational lensing: weak – methods: data analysis – cosmology: observations.

## 1 INTRODUCTION

In the late 1990s, astronomers observing distant Type 1a supernovae made the astonishing discovery that the expansion of the Universe is accelerating (Riess et al. 1998; Perlmutter et al. 1999). This was contrary to the expectation that the gravitational pull of all the matter in the Universe should cause the expansion rate to decrease over time. The accelerated expansion implies the existence of a new form of energy, dubbed 'dark energy' which, according to the standard cosmological model, makes up around 68 per cent of the energy density of the Universe (Abbott et al. 2019; Planck Collaboration 2020), with the remainder cold dark matter (CDM) and ordinary (baryonic) matter.

Within this 'concordance' model, known as ΛCDM, dark energy is a constant energy density filling space homogeneously and resulting in a cosmological constant, Λ, or 'vacuum energy'. Extensions to the concordance model include, most notably, 'quintessence', in which dark energy is a dynamic quantity with energy density that varies in space and time and with an equation of state parametrized by $w(z)$, the pressure to energy density ratio. Precision measurements of $w$ can help distinguish between a cosmological constant, in which $w = -1$ and quintessence, where $w(z) \geq -1$. For a universe with accelerated expansion, $w < -1/3$.

Alternative theories have been posed that do not postulate an additional energy density to explain the accelerated expansion (i.e. non-standard cosmological models), for example, modifications to general relatively at cosmological scales (e.g. Joyce, Lombriser & Schmidt 2016), although results from gravitational wave astronomy have made this less popular (Lombriser & Lima 2017).

One of the most promising probes of the cosmological model is weak gravitational lensing (see, for example Albrecht et al. 2006), in which light emitted by distant galaxies is coherently distorted as it travels through the intervening large-scale structure of the Universe towards the observer. Gravitational lensing is sensitive to distance ratios between the source, lens and observer, as well as the evolution of the matter power spectrum (Bartelmann & Schneider 2001). Given that dark energy affects the growth of structure (on large scales, competing with gravity), the statistics of the distortions to galaxy shapes (or 'shear field'), together with the source and lens redshift information, thus puts constraints on the dark energy equation of state (Hu 1999; Van Waerbeke & Mellier 2003).

Weak lensing is a primary science driver for several Stage IV surveys, including the European Space Agency's *Euclid*[1] satellite (Laureijs et al. 2011; Amendola et al. 2013), launched on 2023 July 1, and the ground-based Legacy Survey of Space and Time[2] (LSST) at the Vera C. Rubin Observatory in Chile (LSST Dark Energy Science Collaboration 2012), with expected first light in August 2024. In addition, the Chinese Survey Space Telescope (CSST or Xuntian; e.g. Gong et al. 2019) and NASA's Nancy Grace Roman Space

---

⋆ E-mail: lv18675@essex.ac.uk

[1] https://www.euclid-ec.org/
[2] https://rubinobservatory.org/

Telescope[3] (formerly the Wide-Field Infra-Red Survey Telescope or WFIRST; Spergel et al. 2015), are due for launch in 2024 and 2027, respectively. In order to realize the full potential of these surveys, weak lensing systematics must be understood and controlled. This is a highly non-trivial task that involves: modelling effects associated with the telescope's optical system and detector (Cypriano et al. 2010; Voigt et al. 2012; Cropper et al. 2016; Eriksen & Hoekstra 2018); overcoming model-fitting bias (Voigt & Bridle 2010), noise bias (Kacprzak et al. 2012; Refregier et al. 2012), and their interaction (Kacprzak et al. 2014); accurate measurements of galaxy photometric redshifts; and the appropriate treatment of intrinsic allignments (e.g. Joachimi & Bridle 2010), selection effects (e.g. Jarvis et al. 2016), and object blending (Samuroff et al. 2018; MacCrann et al. 2021).

Much progress has been made within the weak lensing community to improve shear estimation methods. In particular, the Shear TEsting Programme (STEP; Heymans et al. 2006; Massey et al. 2007) and GRavitational lEnsing Accuracy Testing (GREAT) Challenges (Bridle et al. 2010; Kitching et al. 2012; Mandelbaum et al. 2015) put the state-of-the art methods to the test. Following the GREAT3 challenge, methods including Bayesian Fourier Domain (BFD; Bernstein et al. 2016) and METACALIBRATION (Huff & Mandelbaum 2017; Sheldon & Huff 2017) have been developed that reduce biases below the levels required in next-generation surveys, assuming the noise and PSF are sufficiently well understood.

In recent years, machine learning (ML) has also been applied to shear measurement, utilizing feed-forward artificial neural networks (ANNs) with properties measured from the galaxy images (e.g. ellipticities, fluxes, sizes) as input features (Gruen et al. 2010; Tewes et al. 2019; Pujol et al. 2020), or, alternatively, the galaxy images themselves (e.g. Ribli, Dobos & Csabai 2019; Springer et al. 2020; Zhang et al. 2023). ML has long been used to estimate galaxy photometric redshifts (Collister & Lahav 2004; Brescia et al. 2021, and references therein) and is increasingly being utilized in cosmology (e.g. Fluri et al. 2022) and other areas of astronomy, for example to identify transients (Lopez Portilla et al. 2020; Ayyar et al. 2022).

In this paper, we look at using a convolutional neural network (CNN) for precision weak lensing measurements in a *Euclid*-like survey. CNNs are used to capture information from pixellated input images, extracting features from the images and mapping them to output values or target labels. These models are often applied to classification problems, but can also be used in regression tasks, as in this work. In supervised ML, the parameters of the network are learnt from labelled 'training' data, and then the model's performance is assessed using 'test data', usually a random sub-set of the training data for which the predicted and true labels can be compared. Common to all ML methods is the dependence on the fidelity of the training set; in this application, if the training set does not accurately represent observed galaxies in the Universe then, when it is deployed on survey images, there will be a bias, known in the ML community as 'domain bias' or 'data set shift'. ANNs can fail dramatically when applied to out-of-distribution data.

Studies quantifying the distributions of galaxy properties, such as morphology, bulge fraction, colour, surface brightness, and their correlations are numerous (e.g. Conselice 2006; Calvi et al. 2012; Zhang & Yang 2019); however, there is a limit to how well these distributions can be measured (e.g. Davari, Ho & Peng 2016) and they will depend on the galaxy environment (D'Eugenio et al. 2015; Chen, Hwang & Ko 2016) and specific survey parameters, including, for example, the survey depth, redshift bins, colour bands and selection criteria (Lee, Chary & Wright 2018; Häußler et al. 2022). Thus, it is important to understand the sensitivity of a particular ML model to differences between the training set galaxy images and the actual distribution of observed galaxy morphologies in the Universe.

Shape estimation is also sensitive to how well the point spread function (PSF; see Section 3.2) can be estimated using stars in the field (Paulin-Henriksson et al. 2008; Bertin 2011; Cropper et al. 2013). Notably, Schmitz et al. (2020) find that propagation of the same modelling errors in the PSF through to galaxy ellipticity estimates is dependent on the specific shape measurement method employed.

In this paper, we build on previous work applying ANNs to shear measurement by investigating the impact of using shifted or out-of-distribution data to train the network. Specifically, we look at the effect on the accuracy of shear estimates from using either an incorrect galaxy population, referred to here as 'population bias', or the wrong PSF model (PSF misestimation), in the training sets.

The paper is organized as follows. In Section 2, we briefly review the effect of gravitational shear on an elliptical source and summarize the shear bias requirements for *Euclid*. In Section 3, we describe the galaxy and PSF models and the simulations used to generate pixellated images on postage stamps. In Section 4, we describe the CNN model architecture and define the shear estimator. Section 5 provides details about the galaxy population used in the training sets. In Section 6, we outline the test set simulations and explain how the shear biases are calculated. In Section 7, we optimize the CNN model. In Sections 8 and 9, we quantify the impact of PSF misestimation and galaxy population bias, respectively. Finally, in Section 10, we discuss the results and future work.

## 2 GRAVITATIONAL SHEAR

### 2.1 The lensed ellipticity

For an elliptical source galaxy with minor to major axis ratio $b/a$ and position angle $\phi$, measured counter-clockwise from the $x$-axis to the major axis, the intrinsic source (i.e. unlensed) complex ellipticity is:

$$e^{\text{int}} = \left(\frac{a-b}{a+b}\right)e^{2i\phi} = e_1^{\text{int}} + ie_2^{\text{int}} \quad (1)$$

where $e_1^{\text{int}}$ ($e_2^{\text{int}}$) is the component of the ellipticity along (at 45° to) the $x$-axis. Defining the complex shear, $\gamma = \gamma_1 + i\gamma_2$, galaxy images are distorted by a Jacobian magnification matrix, $M$, given by

$$M = \begin{pmatrix} 1-\kappa-\gamma_1 & -\gamma_2 \\ -\gamma_2 & 1-\kappa+\gamma_1 \end{pmatrix} \quad (2)$$

such that the observed (i.e. lensed) complex ellipticity, $e^{\text{len}}$, is

$$e^{\text{len}} = \frac{e^{\text{int}} + g}{1 + g^* e^{\text{int}}}, \quad (3)$$

where $g = \gamma/(1-\kappa)$ is the reduced shear and $\kappa$ the convergence (Seitz & Schneider 1997). In the weak lensing regime, $\kappa \ll 1$ and $g \approx \gamma$, so that

$$e_1^{\text{len}} \approx e_1^{\text{int}} + \gamma_1, \; e_2^{\text{len}} \approx e_2^{\text{int}} + \gamma_2. \quad (4)$$

In the standard cosmological model, the universe is homogeneous on large scales, and thus we do not expect any preferential orientation of galaxies on the sky. Averaging over galaxies, we find that the two

---
[3]https://roman.gsfc.nasa.gov/

components of the observed shear are given by

$$\gamma_i^{\text{obs}} \approx \langle e_i^{\text{len}} - e_i^{\text{int}} \rangle = \langle e_i^{\text{len}} \rangle \pm \frac{\sigma_i^{\text{int}}}{\sqrt{n_{\text{gal}}}}, \quad (5)$$

where $\sigma_i^{\text{int}} = \sqrt{\langle (e_i^{\text{int}})^2 \rangle}$, with $i = \{1, 2\}$, is the dispersion of the source ellipticity distribution for component $i$, referred to as 'shape noise', and $n_{\text{gal}}$ is the number of galaxies. Shape noise is $\sim 0.3$ (see, e.g. D'Eugenio et al. 2015; Li et al. 2023), approximately an order of magnitude larger than the shear signal.

In addition to a shear, the Jacobian matrix causes an enlargement of the image such that a galaxy with unlensed area given by $ab$ becomes a lensed galaxy with area $ab/(1 - |\gamma|^2)$.

### 2.2 Shear bias definition and requirements

In practice, measurements of galaxy ellipticities, $\hat{e}_i$, are biased estimates of $e_i^{\text{len}}$. Thus, the shear estimator, $\hat{\gamma}_i = \langle \hat{e}_i \rangle$, is a biased estimate of $\gamma_i^{\text{obs}}$. Following the analysis in STEP (Heymans et al. 2006) and subsequent weak lensing studies, we assume a linear relationship between the estimated shear, $\hat{\gamma}_i$, and true shear, such that:

$$\hat{\gamma}_i = (1 + m_i)\gamma_i + c_i, \quad (6)$$

where $m_i$ and $c_i$ are referred to as the multiplicative and additive biases, respectively.[4] The requirements on $m$ and $c$ depend on the survey parameters, including the survey area, galaxy surface density, and median redshift (Amara & Réfrégier 2008). For *Euclid*, the top-level requirements (i.e. including *all* potential sources of bias, as summarized in Section 1) are $|m_i| < 2 \times 10^{-3}$ and $|c_i| < 2 \times 10^{-4}$. For a comprehensive summary of the various bias contributions to weak-lensing shear estimation with *Euclid* see Cropper et al. (2013). For comparison, we also include in the plots bias requirements for the ground-based Dark Energy Survey (DES; 2013–2019)[5], representing a recent (completed) survey.

## 3 SIMULATING THE PSF-CONVOLVED GALAXY IMAGES

In this section, we describe the galaxy and PSF models and the simulations used to generate PSF-convolved galaxy images on postage stamps.

### 3.1 The galaxy model

Typically, galaxy light distributions are represented by a family of Sérsic profiles (Sersic 1968) with intensity $I(x)$ at position $x$ given by

$$I(x) = I_0 \exp\left\{-k\left[(x - x_0)^T C (x - x_0)\right]^{1/2n_s}\right\}, \quad (7)$$

where $n_s$ is the Sérsic index, $I_0$ is the intensity at the galaxy centre, $x_0$, and the covariance matrix, $C$, given by

$$C = \begin{pmatrix} C_{11} & C_{12} \\ C_{21} & C_{22} \end{pmatrix}, \quad (8)$$

has elements

$$C_{11} = \frac{\cos^2(\phi)}{a^2} + \frac{\sin^2(\phi)}{b^2}, \quad (9)$$

$$C_{12} = C_{21} = \frac{1}{2}\left(\frac{1}{a^2} - \frac{1}{b^2}\right)\sin(2\phi), \quad (10)$$

$$C_{22} = \frac{\sin^2(\phi)}{a^2} + \frac{\cos^2(\phi)}{b^2}, \quad (11)$$

with $a$, $b$, and $\phi$ as defined in Section 2.1.

In this work, we simulate a population of disc galaxies, represented by an exponential profile ($n_s = 1$) and ellipticals, modelled by a de Vaucouleurs profile ($n_s = 4$), with constant ellipticity isophotes. Defining $k = 1.9992n_s - 0.3271$, then for a circular galaxy the half-light radius, $r_h = a = b$ (also known as the effective radius), is the radius enclosing half the total flux.[6] The full-width at half-maximum intensity (FWHM) is related to the half-light radius (for a circular profile) through the equation:

$$\text{FWHM} = 2r_h \left(\frac{\ln 2}{k}\right)^{n_s}. \quad (12)$$

The galaxy half-light radii and ellipticity distributions used in this work are described in Section 5 and summarized in Table 1.

### 3.2 The PSF model

In addition to shot noise, images of astronomical objects are distorted and degraded due to (i) atmospheric seeing (for ground-based missions), (ii) the optical system (iii) telescope pointing stability (iv) image pixellization and (v) detector effects, including charge transfer leaking and inefficiency and radiation damage. Here, we ignore detector effects and model the PSF by a convolution with the source intensity profile. Following Voigt & Bridle (2010) and others (e.g. Ribli, Dobos & Csabai 2019), we use a single Gaussian ($n_s = 0.5$) to model the PSF profile. We simulate *Euclid*-like observations with a 0.1 arcsec pixel scale (Laureijs 2017) and a PSF with FWHM of 0.17 arcsec and ellipticity $\approx 0.022$ ($e_1^{\text{PSF}} = 0.01$ and $e_2^{\text{PSF}} = 0.02$). We assume the PSF is constant across the field of view (FoV) and over time and ignore the effects of colour dependence (see Section 10 for further comments).

### 3.3 Simulating the pixellated images

The process of simulating the PSF-convolved galaxy images on pixellated postage stamps is depicted in the flow diagram in Fig. 1 and follows the procedure used in Voigt & Bridle (2010) and Voigt et al. (2012). Galaxy and PSF images are first simulated on separate grids each $17^2$ pixels in size. Prior to convolution, each pixel is divided into a grid with $n_{\text{bin}}^2$ 'sub-pixels' and the intensity calculated at the centre of each of these sub-pixels. For the disc and elliptical galaxies, we use $n_{\text{bin}} = 3$ and 5, respectively (where a finer grid is used for the de Vaucouleurs profile to take account of the more 'peaky' light profile). Convolution between the galaxy intensity profile and the PSF is performed numerically on this finer grid.[7] Following the convolution, the flux in each pixel of the PSF-convolved galaxy image is found by summing the intensity from each sub-pixel. Finally, the $17^2$ pixels grid is cut down to provide image postage stamps which are $15^2$ pixels in size. We find that the results do not change when we use the same, finer binning ($n_{\text{bin}} = 5$ and 7 for the disc and elliptical galaxies, respectively), or a larger grid (19 by 19 pixels) for the convolution, in both the training and the test sets. We note, however, that the biases are sensitive to these choices when the values differ in the training and test sets (see also a discussion of the 'pixel integration level' in Voigt & Bridle 2010) and would need to be tested before using the CNN to estimate shears from real data. This is beyond the scope of this paper, but we discuss in Section 10 possible further work to address this issue.

---

[4] Also, $\langle \hat{e}_i \rangle = (1 + m_i)\langle e_i^{\text{len}} \rangle + c_i$
[5] https://www.darkenergysurvey.org/

[6] We use the terms 'effective radius', 'half-light radius', and 'size' interchangeably. We note that 'size' is often used to refer to the galaxy 'area' measured using quadrupole moments (see Section 10).
[7] The numerical convolution is performed using signal.convolve2d from the SciPy open-source PYTHON library.

**Table 1.** A summary of the PSF and galaxy parameters used in the training set. The value shown for the galaxy half-light radius is the mode, with minimum and maximum allowed values shown in square brackets. $R(\sigma)$ is a Rayleigh distribution with mode $\sigma$. The subscript in square brackets shows the lower and upper allowed ellipticities. See Sections 3.2 and 5 for further details.

|         | Sérsic index | Half-light radius (arcsec) | Ellipticity | Ellipticals:discs | Size–magnitude relation ($\alpha_r, \beta_r$) |
|---------|--------------|----------------------------|-------------|-------------------|----------------------------------------------|
| PSF     | 0.5          | 0.08372                    | 0.022       | –                 | –                                            |
| Galaxies| 1, 4         | $0.3_{[0.2,\,0.8]}$        | $R(0.25)_{[0,0.7]}$ | 1:4       | $(-0.1286, 2.65)$                            |

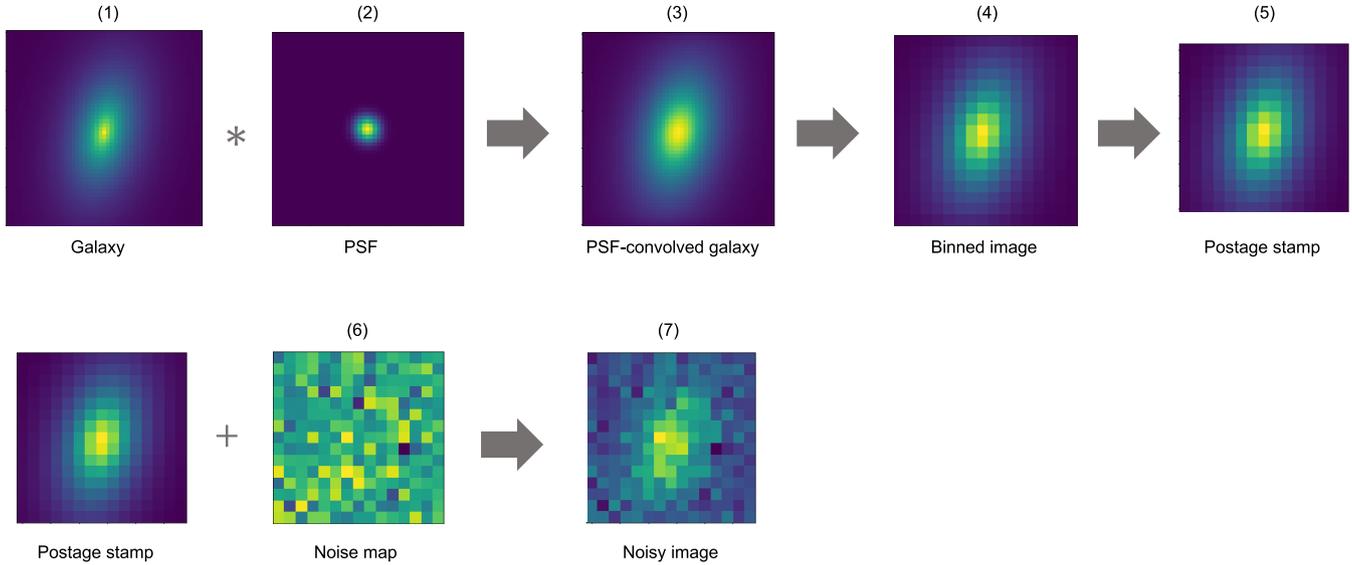

**Figure 1.** Flow diagram showing the steps involved in simulating a noisy PSF-convolved galaxy image on a 15 by 15 pixel postage stamp (see Section 3.3): (1) galaxy image simulated on a 17 by 17 pixel grid with each pixel divided into $n_{\text{bin}}^2$ sub-pixels; (2) PSF simulated on the same 'fine' grid used in step 1; (3) galaxy convolved numerically with the PSF on the fine grid; (4) fine grid binned up by a factor of $n_{\text{bin}}$; (5) binned grid cut down to central 15 by 15 pixels; (6) noise map generated, and (7) added to the postage stamp, creating the final image. Shown for a disc galaxy ($n_s = 1$) and $S/N = 34$.

## 4 THE CNN MODEL

### 4.1 The model architecture

We train two separate ANNs: one to label each galaxy with an $\hat{e}_1$ estimate and another to label each galaxy with an $\hat{e}_2$ estimate. The networks are built using Keras Sequential models, provided by TensorFlow (Abadi et al. 2015) and have the same shallow architecture, shown in Fig. 2.

Input images contain a single PSF-convolved galaxy on a 15 by 15 pixel postage stamp, simulated using the Gaussian PSF described in Section 3.2. In the test sets, we assume a constant PSF which is either the same as the one used in the training sets (Sections 7 and 9), or has a different size or ellipticity to the PSF used in the training sets (Section 8). In practice, the PSF varies with position and time, as well as depending on the spectral energy distribution (SED) of the source. We discuss this further in Section 10.

The first layer in each network is a convolutional layer[8] with $n_{\text{fil}}$ filters (or kernels), each 3 by 3 pixels in size, and with a stride of one. We do not pad the images and therefore the output 'feature maps' are each 13 by 13 pixel grids. The grid values in each feature map, $u_{i,f}$, where $f$ denotes the filter, are found by sliding the kernel across the input image, moving along and then down one pixel at a time, and, at each kernel position, computing

$$u_{i,f}(j,k) = g(w_{i,f} \cdot I(j:j+2, k:k+2) + b_{i,f}), \quad (13)$$

where $j, k \in \{0, 1, \ldots, 12\}$, $w_{i,f}$ are the filter weights, $b_{i,f}$ the bias, and $g$ is the activation function, chosen here to be a Rectified Linear Unit (ReLU; Nair & Hinton 2010), such that $g(y) = \max(0, y)$. The number of fitted parameters in this layer is $10 \times n_{\text{fil}}$ (i.e. $3^2$ weights and one bias for each filter).

The feature maps are then passed through the next layer,[9] which flattens the output from the previous layer, with shape (batch_size, 13, 13, $n_{\text{fil}}$), where batch_size is the number of samples used per gradient update, into a tensor with shape (batch_size, $N$), where $N = 13^2 \times n_{\text{fil}}$.

---

[8]tensorflow.keras.layers.Conv2D
[9]tensorflow.keras.layers.Flatten

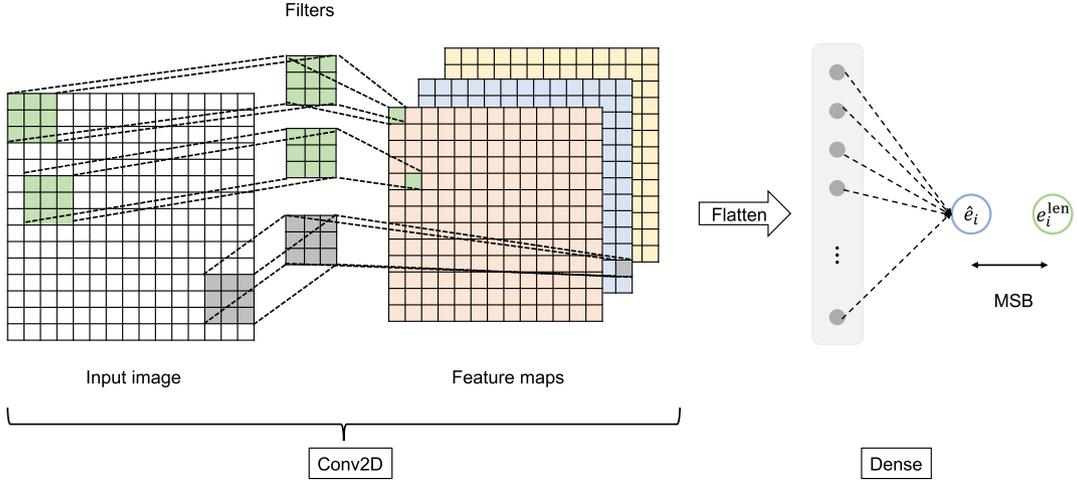

**Figure 2.** Diagram showing the CNN architecture for a single 15 by 15 pixel input image containing a PSF-convolved galaxy. The network is trained on $n_{\text{gal}}$ images using the same PSF for each galaxy, described in Section 3.2. Two different filters are shown for illustration, with 30 filters used in the actual model. $\hat{e}_i$ and $e_i^{\text{len}}$ are the estimated and true lensed galaxy ellipticities, defined in Section 2. See Section 4 for a detailed description of each layer.

The next layer is a dense layer[10] connecting the output values $u_{i,p}$, where $p = \{1, 2, \ldots, N\}$, from the previous (flattened) layer to a single output, which is the ellipticity estimator, $\hat{e}_i$, for each input galaxy image in the batch. We use a hyperbolic tangent for the activation function to ensure the output takes values between $-1$ and 1. The output label, or ellipticity estimate, is thus

$$\hat{e}_i = \tanh z_i, \tag{14}$$

with

$$z_i = \sum_{p=1}^{N} (W_{i,p} u_{i,p} + B_i), \tag{15}$$

where $W_{i,p}$ and $B_i$ are the weights and the bias connecting the dense layer to the output. We do not include any pooling or dropout layers in the network. The fiducial CNN model hyperparameters are shown in Table 2.

### 4.2 The loss function

For the loss function, in order to mitigate the effects of noise bias[11], we follow Gruen et al. (2010) and Tewes et al. (2019) by adopting a 'mean square bias' (MSB), given by:

$$\text{MSB} = \frac{1}{n_{\text{gal}}} \sum_{n=1}^{n_{\text{gal}}} \left[ \frac{1}{n_{\text{real}}} \sum_{m=1}^{n_{\text{real}}} \left( e_i^{\text{len}} - \hat{e}_i \right) \right]^2, \tag{16}$$

where $n_{\text{gal}}$ is the number of distinct, noise-free galaxy images in the training set, and $n_{\text{real}}$ is the number of noisy realizations of each of these images. Thus, the total number of noisy galaxy images in the training set is $n_{\text{gal}} \times n_{\text{real}}$. We use a batch_size = $n_{\text{real}}$.[12]

---

[10] tensorflow.keras.layers.Dense

[11] Because ellipticity ($e_i$) is not a linear sum over pixel intensities, for noisy galaxy images, the ordinary least squares (OLS) estimators (in which the loss function is the mean square error between the true and predicted $e_i$ values) are biased, i.e. the expected value of the error term in the regression is not zero.

[12] Note that for $n_{\text{real}} = 1$, the loss function reduces to the mean square error (MSE) and the batch size to 1.

### 4.3 The shear estimator

We build multiple CNN models, where each model is trained on a different random set of noisy galaxy images (drawn from the same underlying distribution, described in Section 5). In addition, each CNN model is given a different seed to start the training process. We note that we do not apply any shear to the galaxies in the training sets.

A given test set (representing observed galaxies, see Section 6) is passed through a 'committee' of trained CNN models, with each model providing a shear estimate, $\hat{\gamma}_i^{\text{cnn}}$, which is the mean over all predicted ellipticities in the test set, i.e.

$$\hat{\gamma}_i^{\text{cnn}} = \langle \hat{e}_i \rangle. \tag{17}$$

For $n_{\text{cnn}}$ models in the committee, our shear estimator, $\hat{\gamma}_i$, is given by

$$\hat{\gamma}_i = \frac{1}{n_{\text{cnn}}} \sum_{q=1}^{n_{\text{cnn}}} \hat{\gamma}_{i,q}^{\text{cnn}} \pm \frac{s_i^{\text{cnn}}}{\sqrt{n_{\text{cnn}}}}, \tag{18}$$

where $s_i^{\text{cnn}}$ is the unbiased sample standard deviation over the shear estimates, $\hat{\gamma}_{i,q}^{\text{cnn}}$, for a given test set. By the Central Limit Theorem, provided that $n_{\text{cnn}} \gtrsim 30$, the distribution of the shear estimator is approximately normal. In practice, we use $n_{\text{cnn}} = 35$.

## 5 THE TRAINING SETS

In this section, we describe the properties of the training set galaxies (which are used to fit the CNN model weights and biases given in equations (13) and 15). We simulate a population of galaxy images with 20 per cent de Vaucouleurs and 80 per cent exponential profiles, chosen to approximately represent the observed proportion of elliptical galaxies to galaxies containing discs.

We adopt a power-law distribution for the number density of galaxies as a function of apparent magnitude, $m_{\text{AB}}$, as follows:

$$p(m_{\text{AB}}) = (m_{\text{AB,u}} - m_{\text{AB,l}})(\alpha_m + 1) m_{\text{AB}}^{\alpha_m} + m_{\text{AB,l}}, \tag{19}$$

where we use $\alpha_m = 0.36$ (Hoekstra, Viola & Herbonnet 2017) and magnitudes in the range $m_{\text{AB,l}} = 20.5$ and $m_{\text{AB,u}} = 24.5$, with the upper magnitude chosen to correspond to the performance requirement of *Euclid*'s visible imager (*Euclid*-VIS) for extended sources of order 0.3 arcsec.

It is known that the morphological properties of galaxies (e.g. size, ellipticity, surface brightness profile) and their apparent magnitudes are correlated (see, e.g. Euclid Collaboration: Martinet et al. 2019). In this paper, we consider the size-magnitude correlation, but a full study including correlations between several galaxy properties is beyond the scope of this work.

We use a relationship between galaxy apparent magnitude and size from Hoekstra, Viola & Herbonnet (2017) (see their fig. 2), such that:

$$\langle \log_{10} r_h \rangle = \alpha_r m_{AB} + \beta_r \quad (20)$$

and

$$\sigma_{\log_{10} r_h} = \alpha_\sigma m_{AB} + \beta_\sigma, \quad (21)$$

with $\alpha_r = -0.12857$, $\beta_r = 2.65$, $\alpha_\sigma = -0.0166$, and $\beta_\sigma = 0.56$ for $r_h$ measured in arcsec. We draw half-light radii assuming a normal distribution with

$$\log_{10} r_h \sim N\left(\langle \log_{10} r_h \rangle, \sigma^2_{\log_{10} r_h}\right). \quad (22)$$

We make cuts on the pre-lensed galaxy size such that $0.2 \leq (r_h/\text{arcsec}) \leq 0.8$. We comment on the lower cut-off in Section 10. A scatter plot showing the relationship between the galaxy size and apparent magnitude before and after cuts on signal-to-noise (S/N; see below) is shown in Fig. 3.

The unlensed galaxy ellipticity is drawn from a truncated Rayleigh distribution with mode $e^{int} = 0.25$ and a maximum value $e^{int} = 0.7$. The major and minor axes lengths are calculated using $ab = r_h^2$ and $e = (a-b)/(a+b)$, as in equation (1). Histograms showing the distributions of galaxy apparent magnitude, size and ellipticity are shown in Fig. 4.

The remaining four parameters in equation (7) defining the galaxy intensity profile are orientation, peak intensity, and centroid. The galaxy orientation is drawn from a uniform distribution with $\{\phi \in \mathbb{R} : 0 \leq \phi < \pi\}$. The galaxy centroid position is randomized uniformly within the central pixel of the postage stamp image. The peak intensity, $I_0$, is related to the flux via the equation:

$$I_0 = \frac{F}{2\pi n_s k^{-2n_s} r_h^2 \Gamma(2n_s)}, \quad (23)$$

where $\Gamma$ is the gamma function and the flux, $F$, is given by:

$$F = F_0 10^{-0.4 m_{AB}}. \quad (24)$$

Galaxies are simulated on pixellated grids and convolved with the PSF model as described in Section 3. The galaxy and PSF model parameters used in the training sets are summarized in Table 1. We approximate the finite number of photons arriving on the detector by adding uncorrelated noise to each pixel in the postage stamp (containing the PSF-convolved image), drawn from a Gaussian distribution given by $G(0, \sigma_n^2)$, with $\sigma_n$ constant across pixels. We note, however, that undetected galaxies act as a source of correlated noise (see Euclid Collaboration: Martinet et al. 2019).

The signal-to-noise ratio is defined as:

$$S/N = \frac{\sqrt{\sum I(x)^2}}{\sigma_n}, \quad (25)$$

where the sum is taken over all the image pixels in the postage stamp. We choose the ratio $F_0/\sigma_n$ to give a signal-to-noise distribution with mode ~11 (see Fig. 4). We remove all galaxies with $S/N < 10$ or $S/N > 100$. Examples of noisy PSF-convolved galaxy images on postage stamps are shown in Fig. 5 for a range of S/N values.

## 6 THE TEST SETS

Galaxy test sets are simulated to represent observed galaxies. We first simulate test sets with galaxies drawn from the same distribution as the training set galaxies and with the same PSF (see Sections 5). We then look at the effects of using either the wrong PSF model (Section 8) or an incorrect galaxy population (Section 9), in the training sets. In practice, we change the parameters used in the test sets to be different from those in the training sets. We apply the same size and ellipticity restrictions to the pre-sheared test set galaxies as used in the training sets, except that the upper allowed ellipticity in the test sets is 0.6, as opposed to 0.7 in the training sets. We note that we do not consider selection biases (see, for example, Jarvis et al. 2016); all galaxies are simulated on individual postage stamps and included in the sample if they meet the signal-to-noise criteria (see Section 5). Cuts on galaxy size and ellipticity are made prior to simulating the images. We do not investigate the impact of different S/N, size or ellipticity cuts.

We make the simplifying assumption that all galaxies in a test set (i.e. across the FoV) are subjected to the same constant shear. 25 test sets are generated using the following sets of values for each component of the shear: $\gamma_1 = \{-0.025, -0.01, 0, 0.01, 0.025\}$ and $\gamma_2 = \{-0.05, -0.015, 0, 0.015, 0.05\}$. We use all $(\gamma_1, \gamma_2)$ combinations from these shear sets.

Each test set is a different random realization of galaxies and noise maps. For each pre-sheared galaxy in a test set, a second is generated that is orthogonal to the first. This removes the shape noise[13], described in Section 2 (see also Massey et al. 2007), so that, for a perfect shear measurement method, the estimated ellipticity, $\hat{e}_i$, averaged over all galaxies in the test set, will be equal to the shear $\gamma_i$. As such, we need only generate enough galaxies in each test set to fairly represent the distribution of galaxy shapes (e.g. morphologies, orientations, sizes and ellipticities) and to reduce the uncertainty from noise to the required level (i.e. to reach the required precision). In practice, for noise-free images (which we use to optimize the CNN model hyperparameters and choose the number of galaxies required in the training set), we use $4 \times 10^4$ rotated (or 'matched') pairs of galaxies in each test set (i.e. $8 \times 10^4$ galaxies). For noisy images, we use $4 \times 10^5$ rotated pairs (i.e. $8 \times 10^5$ galaxies). We note that, for the unsheared test set and in the absence of pixellization, the pre-PSF convolved galaxies in a pair are identical apart from the 90° rotation.

For each one of the 25 test sets, we obtain shear estimates $\hat{\gamma}_1$ and $\hat{\gamma}_2$, which are the mean estimates from the committee of trained CNN models (see equation (18)). Multiplicative and additive biases are then calculated using ordinary least squares regression (see equation (6) and Section 7).

We note that the CNN models are built using unsheared training set galaxies (see Section 4). The test sets containing sheared galaxies are thus drawn from a different distribution to the galaxies in the training sets, even for the same galaxy parameter distributions and PSF.

## 7 OPTIMIZING THE CNN

In this section, we find the number of training set galaxies ($n_{gal}$) and noise realizations per galaxy ($n_{real}$) required in the training set in order to reduce the multiplicative and additive biases to an acceptable level. We also optimize the CNN model hyperparameters; specifically, the number of epochs used in the training process and the number of filters in the first layer of the network.

---

[13]Shape noise contributes to the dispersion in the additive bias.

**Table 2.** CNN fiducial hyperparameters. Note that we use $n_{\rm real} = 1$ (500) for noise-free (noisy) training set images. See Sections 4 and 7 for further details.

| Number of filters $n_{\rm fil}$ | Filter width (pixels) | Filter stride (pixels) | Number of epochs | Batch size | Learning rate | Number of galaxies $n_{\rm gal}$ |
|---|---|---|---|---|---|---|
| 30 | 3 | 1 | 100 | $n_{\rm real}$ | 0.001 | $10^5$ |

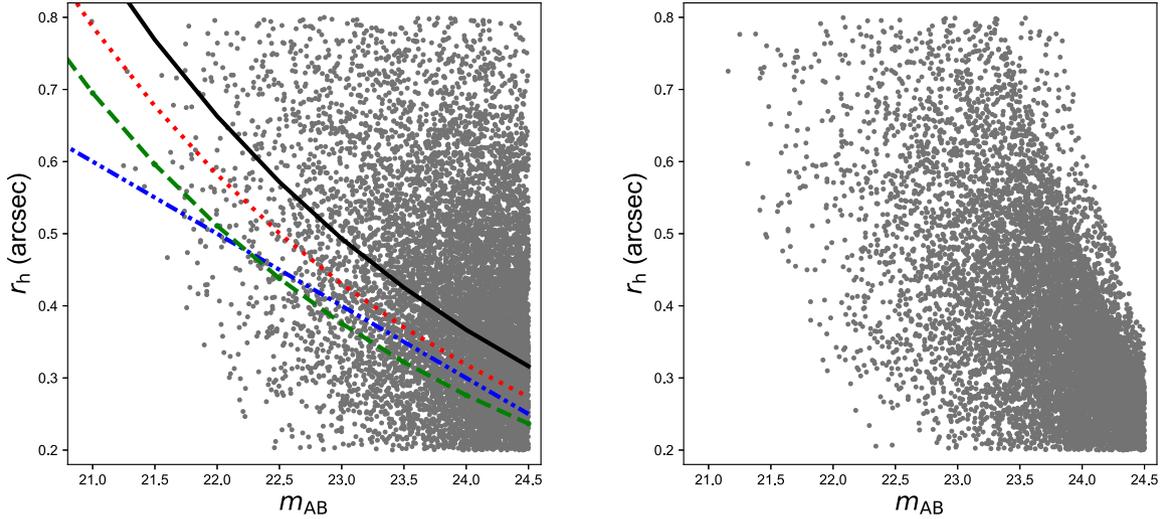

**Figure 3.** Relationship between half-light radius and apparent magnitude before (left) and after (right) cuts on signal-to-noise ($S/N \geq 10$; both plots exclude $S/N > 100$). Results shown for $10^4$ galaxies using the parameters adopted for the training sets (see Section 5 for details). Each point represents a galaxy. Curves in the left-hand plot show the size–magnitude relation given in equation (20) using the slope adopted in the training set (black solid; $\alpha_{\rm r} = -0.1286$) and with a 2 per cent (red dotted; $\alpha_{\rm r} = -0.1311$) and 4 per cent (green dashed; $\alpha_{\rm r} = -0.1337$) steeper slope. $\beta_{\rm r} = 2.65$ is fixed. The blue dash–dotted curve shows the size–magnitude relation adopted in Euclid Collaboration: Martinet et al. (2019, see their fig. 1), corresponding approximately to a 3 per cent steeper slope. See Section 9 for a discussion of the shear biases arising from using an incorrect slope for the galaxy size–magnitude relation in the training set.

For this, we use the same galaxy distribution in the training and test sets, with parameters described in Section 5, and with the correct PSF model (see Sections 8 and 9 for the biases when we use either the incorrect PSF model, or a different galaxy distribution, respectively, in the test sets). Using noise-free images in both the training and test sets, we show in Fig. 6 the dependence of the biases on $n_{\rm gal}$, as well as on the number of epochs and filters. We find that, for the model architecture we use here, we require $\gtrsim 10^5$ galaxies to reduce the biases below those required for a *Euclid*-like survey. We find that the biases flatten for epochs $\gtrsim 100$ and number of filters $\gtrsim 20$. In practice, we use $n_{\rm gal} = 10^5$ to train the model, with 30 filters and 100 epochs (see Table 2 for a summary of the fiducial CNN model hyperparameters). We note that, for the model architecture and galaxy population adopted in this study, the multiplicative biases flatten for $n_{\rm gal} \gtrsim 10^5$ and that future work should explore methods to reduce the biases further, for example by using a deeper network.

Finally, we simulate noisy galaxies using the $S/N$ distribution shown in Fig. 4, with $S/N \geq 10$. In Fig. 7, we show the errors on the estimated shears from each individual test set, as a function of the true input shears, for different values of $n_{\rm real}$. Multiplicative and additive biases are calculated by fitting the linear model given in equation (6) to the shear estimates ($\hat{\gamma}_i$), calculated using equation (18). Results showing the dependence of $m$ and $c$ on the number of noise realizations per galaxy are plotted in Fig. 6. Notably, for $n_{\rm real} = 1$, the MSB loss function, given in equation (16) and described in Section 4.2, reduces to the MSE loss function. We see that, with noisy images, the biases are high when we use the

MSE loss function ($n_{\rm real} = 1$), with $|m_i| > 0.1$ (see also, e.g. Kacprzak et al. 2012; Refregier et al. 2012, for the noise bias levels found in other studies using the MSE loss function). In Fig. 6, we find that we need $\gtrsim 500$ noise realizations per training set galaxy in order to reduce the noise bias by approximately two orders of magnitude, reaching the required levels and consistent with the biases found in noise-free images. We note that each trained CNN takes $<0.05$ ms to make an ellipticity prediction and thus a committee of 35 models provides a shear estimate per galaxy image in $<1.75$ ms.

In Sections 8 and 9, we use CNN models built using $n_{\rm gal} = 10^5$ and $n_{\rm real} = 500$, which is sufficient to explore the potential impact of PSF misestimation and galaxy population bias. However, as discussed in Section 10, in future work, the shear measurement biases will need to be reduced even further below the top-level requirements to allow for additional sources of systematics (e.g. Cropper et al. 2013).

## 8 PSF MISESTIMATION BIAS

In this section, we quantify the biases arising from an inaccurate modelling of the PSF in the training sets. Specifically, we consider the impact on the multiplicative and additive biases when we use either an incorrect PSF size, or an incorrect component of the PSF ellipticity, parameterised by

$$m_i = \beta_{0,i} + \beta_{1,i} \left( \frac{\delta r_{\rm h}^{\rm PSF}}{r_{\rm h}^{\rm PSF}} \right) \qquad (26)$$

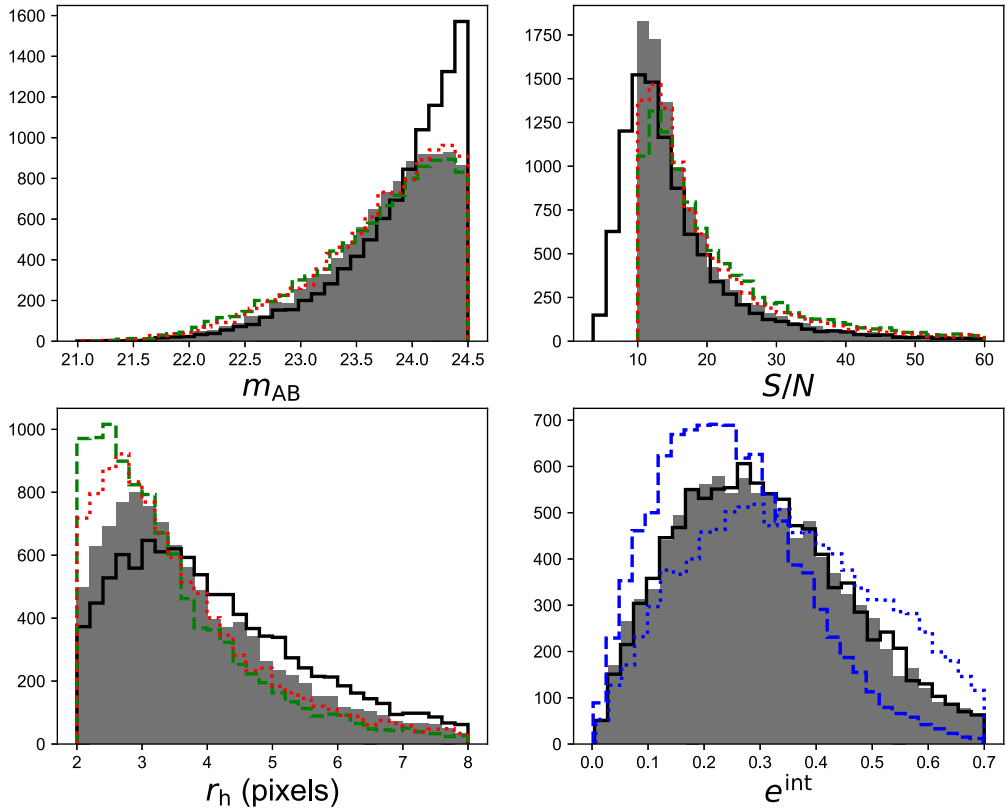

**Figure 4.** Histograms showing the galaxy apparent magnitude (top left), *S/N* (top right), half-light radius (bottom left), and intrinsic ellipticity (bottom right) distributions. Grey shaded areas show the distributions used in the training sets (see Section 5; $10 \leq S/N \leq 100$). Black solid lines show the distributions before the lower signal-to-noise cut. Green dashed and red dotted lines (see Fig. 3 caption for details) show the magnitude, *S/N*, and size distributions used in the test sets to investigate the impact of a shift in the size–magnitude relation (see Section 9). Similarly, the blue lines in the bottom right plot show the impact of an incorrect ellipticity distribution in the training set; specifically, dashed (dotted) lines are for a peak $e^{\mathrm{int}}$ of 0.2 (0.3) in the test sets (with 0.25 used in the training sets; see Table 1). Histograms are shown for $10^4$ galaxies. Note that $\sim 1$ per cent of galaxies have *S/N* values above the upper limit shown on the *x*-axis in the top right plot. The correlation between apparent magnitude and size is shown in Fig. 3. We do not include any correlation between the galaxy's ellipticity and size or magnitude.

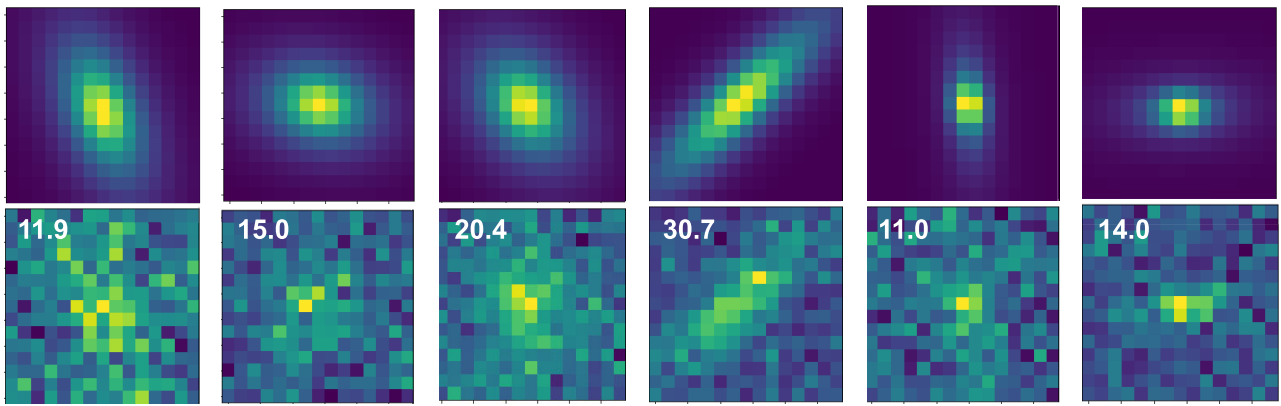

**Figure 5.** Example noisy (bottom) and corresponding noise-free (top) PSF-convolved galaxy images on 15 by 15 pixel postage stamps. The type ($n_{\mathrm{s}}$), size ($r_{\mathrm{h}}$ in pixels), and ellipticity ($e^{\mathrm{int}}$) of each galaxy (from left to right) are, respectively, as follows: (1, 3.7, 0.35); (1, 3.4, 0.17); (1, 3.7, 0.17); (1, 3.8, 0.64); (4, 3.0, 0.58); (4, 4.2, 0.39). The signal-to-noise ratio (see equation (25)) is shown in the top left-hand corner of each noisy image.

and

$$c_i = \alpha_{0,i} + \alpha_{1,i}\delta e_i^{\mathrm{PSF}}, \quad (27)$$

where $r_{\mathrm{h}}^{\mathrm{PSF}}$ is the PSF size used in the test sets and $\delta r_{\mathrm{h}}^{\mathrm{PSF}}$ ($\delta e_i^{\mathrm{PSF}}$) is the difference between the PSF size (ellipticity component) used in the training and test sets. In each case, a positive difference corresponds to a larger value in the test sets (i.e the 'true' value) than in the training sets.

We obtain shear estimates by running the test sets through the CNN models built as described in Section 7, using $10^5$ galaxies and with the PSF model and galaxy distributions described in Sections 3.2 and 5, respectively. Results are obtained for both noisy (using CNN

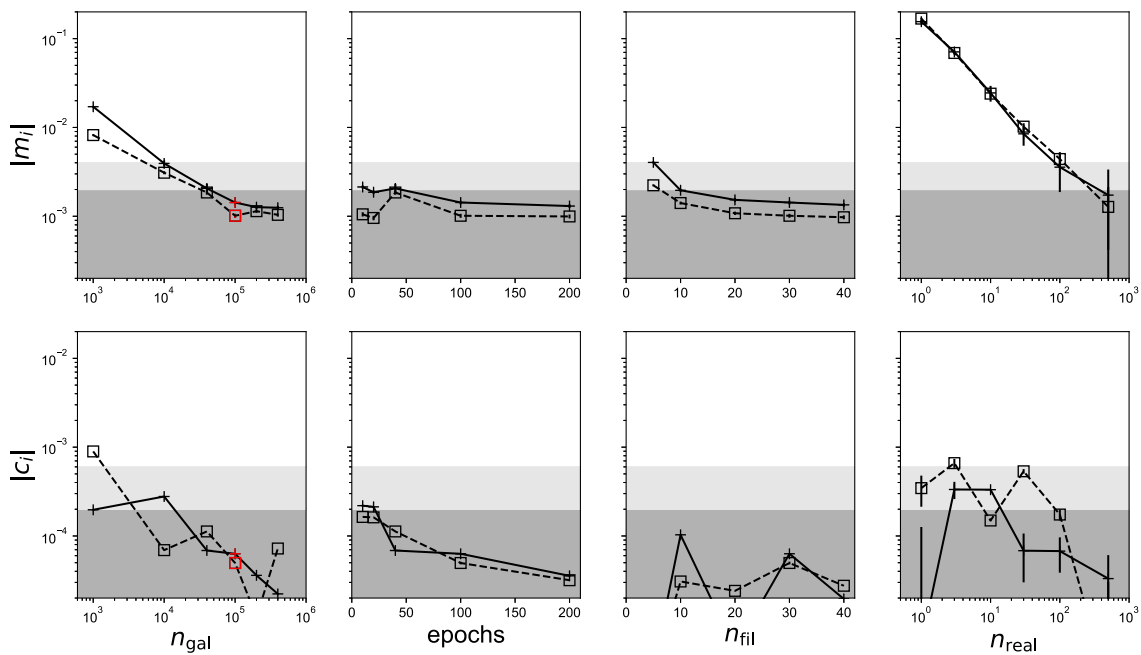

**Figure 6.** Absolute values for the multiplicative $m_i$ (upper panels) and additive $c_i$ (lower panels) biases for $i = 1$ (crosses; solid) and $i = 2$ (open squares; dashed). Plots are shown (from left to right) as a function of the number of galaxies used in the training set ($n_{\rm gal}$), number of epochs used to train the network, number of filters ($n_{\rm fil}$), and number of noisy realizations used per galaxy ($n_{\rm real}$) in the training sets. The first three plots are shown for noise-free simulations. The reference biases, shown in red in the left-most plots are: $m_{1,\,\rm ref} = 1.4 \times 10^{-3}$, $m_{2,\,\rm ref} = 1.0 \times 10^{-3}$, $c_{1,\,\rm ref} = 6.3 \times 10^{-5}$, and $c_{2,\,\rm ref} = 5.0 \times 10^{-5}$; see Section 8). The fiducial hyperparameters used are shown in Table 2. The dark shaded region depicts the bias requirements for *Euclid* and the lighter shaded region for recent surveys, such as DES (see Section 2.2).

models trained on 500 noise realizations per galaxy) and noise-free images.

In Fig. 8, we plot the increase in $m$ and $c$ ($\Delta m = m_i - m_{i,\,\rm ref}$ and $\Delta c = c_i - c_{i,\,\rm ref}$, respectively) relative to the biases we obtain in the fiducial setting for noise-free images ($m_{i,\,\rm ref}$, $c_{i,\,\rm ref}$; see Fig. 6), with only one parameter at a time offset from the values used in the training sets. We plot relative biases in order to remove any offsets in Fig. 8 and subsequent plots (see Section 9), given that the noise-free biases we obtain are not zero, even when we use the same PSF in the training and test sets. We note that $\Delta m_i \approx m_{i,\,\rm cal}$ and $\Delta c_i \approx c_{i,\,\rm cal}$, where $m_{i,\,\rm cal} = \Delta m_i/(1 + m_{i,\,\rm ref})$ and $c_{i,\,\rm cal} = \Delta c_i/(1 + m_{i,\,\rm ref})$, are the biases that would be obtained if we calibrated the estimated shears using the reference biases. In practice, before using on survey data, we would expect to improve the CNN model (in addition to addressing other issues, outlined in Section 10) to reduce $m_{i,\,\rm ref}$ and $c_{i,\,\rm ref}$ to, effectively, zero, thereby removing the need for any calibration.

We find that, as expected (e.g. Heymans et al. 2006), an error in a component of the PSF ellipticity affects the additive shear bias, with an error in $e_1^{\rm PSF}$ increasing $c_1$, but having negligible effect on $c_2$, and vice versa for $e_2^{\rm PSF}$. The multiplicative biases are not significantly affected by an incorrect $e_i^{\rm PSF}$ in the training set and are consistent between noisy and noise-free simulations within the error bars.

For the additive biases, we find that $\Delta c_i$ can be kept below the *Euclid* upper limit of $2 \times 10^{-4}$ if the error in either component of the ellipticity meets the requirement $|\delta e_i^{\rm PSF}| \lesssim 10^{-3}$. We find some deviation between the additive biases obtained for noisy and noise-free simulations for $|\delta e_i^{\rm PSF}| \gtrsim 10^{-3}$, with a stronger dependence on the PSF ellipticity misestimation for noisy images. We fit a regression line to the additive biases for both the noise-free and noisy images for $|\delta e_i^{\rm PSF}| \leq 2 \times 10^{-3}$. We find $\alpha_{1,1} = 0.12$, $\alpha_{1,2} = 0.11$ ($\alpha_{0,1} = -2.2 \times 10^{-5}$, $\alpha_{0,2} = -1.4 \times 10^{-5}$) for the noise-free images and $\alpha_{1,1} = \alpha_{1,2} = 0.17$ ($\alpha_{0,1} = -3.7 \times 10^{-5}$, $\alpha_{0,2} = -5.1 \times 10^{-5}$) for the noisy images.

We now consider the impact of an incorrect PSF size. We offset the true PSF size (used in the test set) by an amount $\delta r_{\rm h}^{\rm PSF}$ from the half-light radius used in the training set, given in Table 1. We plot the biases on the true shear in Fig. 8 as a function of $\delta r_{\rm h}^{\rm PSF}/r_{\rm h}^{\rm PSF}$, where $r_{\rm h}^{\rm PSF}$ is the PSF size in the test sets. The results are consistent between the noisy and noise-free simulations within the error bars, though we note that the biases found for $m_1$ appear to be affected more by the presence of noise than those for $m_2$, with $m_1$ consistently lower in the noisy, as compared to the noise-free, simulations. We find that there is a significant impact on the multiplicative biases, with $\Delta m_i$ rising above the *Euclid* requirements for $|\delta r_{\rm h}^{\rm PSF}|/r_{\rm h}^{\rm PSF} \gtrsim 5 \times 10^{-3}$. We fit regression lines to the mean of $\Delta m_1$ and $\Delta m_2$ in the noisy and noise-free simulations and find that $\beta_1 = \langle \beta_{1,i} \rangle_{i \in \{1,2\}} = -0.24$ in both cases. We measure $\beta_0 = \langle \beta_{0,i} \rangle_{i \in \{1,2\}} = -6.7 \times 10^{-4}$ and $5.8 \times 10^{-5}$ in the noisy and noise-free simulations, respectively. The additive biases are relatively unaffected.

The results presented here quantify the sensitivity of the CNN model to inaccuracies in the training images, specifically as a result of an incorrect PSF. We compare our results to the *Euclid* requirements on the tolerated root mean square (RMS) errors in the PSF model parameters in Section 10.

## 9 GALAXY POPULATION BIAS

In this section, we consider contributions to the galaxy population bias (caused by differences between the galaxy populations in the training and test sets), arising from two distinct effects: (i) incorrect parameter values used to describe either the galaxy ellipticity distribution, size–magnitude relation, or ratio of galaxy types, referred to here as 'galaxy distribution bias'; and (ii) incorrect or insufficient modelling of galaxy light intensity profiles, which we call 'morphology bias' (see also 'model-fitting bias', e.g. Voigt & Bridle 2010).

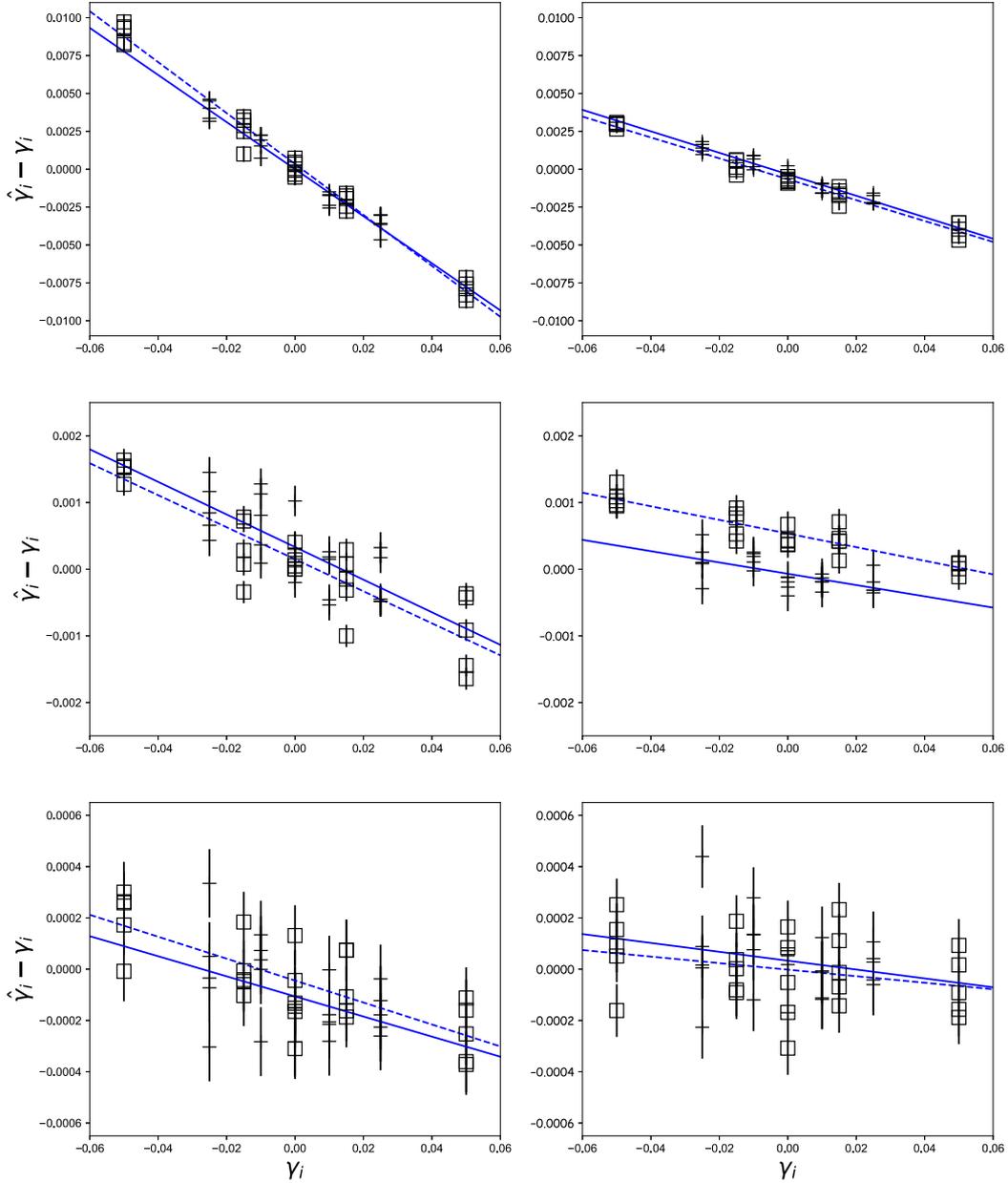

**Figure 7.** Plots showing the differences between the estimated ($\hat{\gamma}_i$) and true ($\gamma_i$) shears for each test set as a function of the true input shears. Crosses (squares) are for the first (second) components of the shear and blue solid (dashed) lines the OLS regression fits. Results are shown for noisy images with an increasing number of noise realizations per galaxy: $n_{\rm real} = 1$ (top left), $n_{\rm real} = 3$ (top right), $n_{\rm real} = 10$ (middle left), $n_{\rm real} = 30$ (middle right), $n_{\rm real} = 100$ (bottom left), and $n_{\rm real} = 500$ (bottom right).

### 9.1 Distribution bias

Here, we introduce shifts in the distributions describing the galaxy populations used in the test sets, as compared to those used in the training sets (see Table 1 for the parameter values used in the training sets). Specifically, we consider, separately, the effects of using: (i) a shifted ellipticity distribution, in which the mode of the Rayleigh distribution is offset from the value used in the training set, but with the same upper and lower bounds; (ii) a different slope for the size–magnitude relation, given by $\alpha_{\rm r}$ in equation (20); and (iii) a different ellipticals to disc galaxies ratio. We use the correct galaxy intensity profiles (i.e. both the training and test sets contain de Vaucouleurs and exponential profiles only), as well as the correct PSF.

We plot the relative biases (see Section 8) in Fig. 9 and find that the results are approximately consistent between the noise-free and noisy images for both the additive and multiplicative biases. We see that a 'data set shift', arising from differences in the distributions describing the galaxy populations in the training and test sets, has a negligible effect on the additive biases. However, there is a significant impact on the multiplicative biases. We find that a shift in the mode of the galaxy ellipticity distribution by more than ∼10 per cent raises the biases above the *Euclid* requirements. The impact from using a different ratio of ellipticals to disc galaxies in the training and test sets is less strong in percentage terms, with a tolerated shift of ∼25–50 per cent of the value adopted in the training set. For the size–magnitude relation, we consider increases in the magnitude of the slope parameter $\alpha_r$ by up to 4 per cent, with curves shown for 2 and 4 per cent steeper[14] slopes in Fig. 3. We keep the intercept, $\beta_r$, constant and thus an increase in the absolute value of the slope corresponds to a shift to smaller galaxies. We plot the shifted distributions used in the test sets in Fig. 4. We find that the biases are unacceptably high for a 2–3 per cent steeper slope in the test sets than used to train the CNN. For reference, we also plot in Fig. 3 the size–magnitude relation adopted in Euclid Collaboration: Martinet et al. (2019), in which the authors quantify the impact of undetected galaxies on shear measurements, corresponding approximately to a 3 per cent steeper slope.

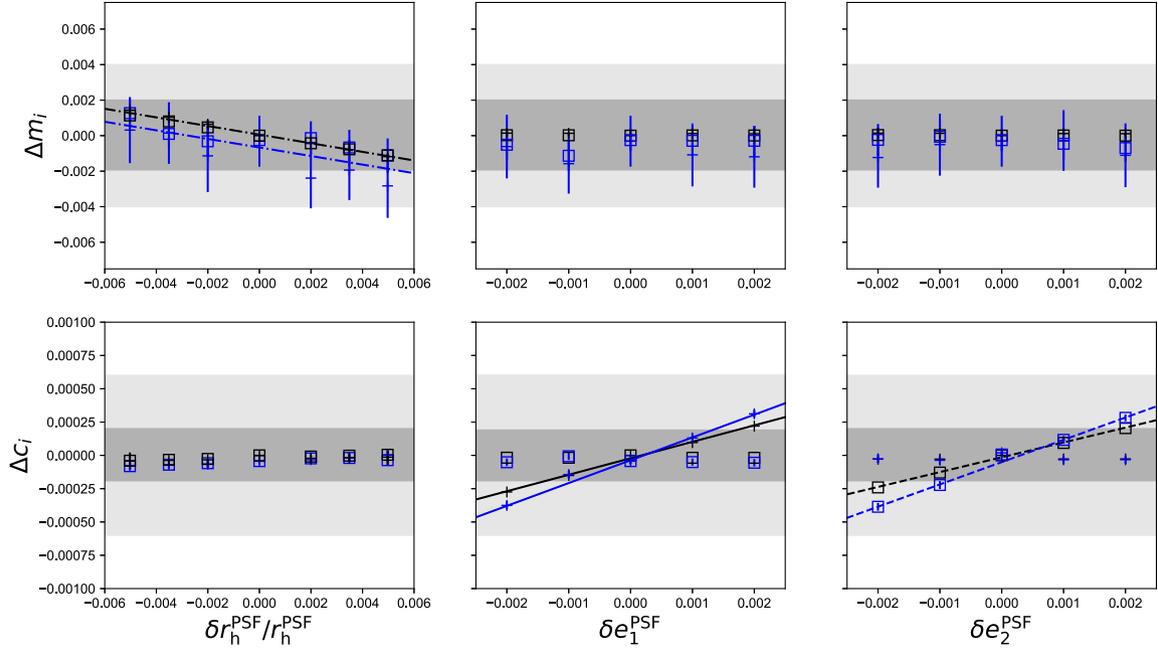

**Figure 8.** Increase in the multiplicative ($\Delta m_i$; upper panels) and additive ($\Delta c_i$; lower panels) biases with respect to the fiducial setting (see Fig. 6 and Section 8) for $i = 1$ (crosses) and 2 (open squares) as a function of the misestimation in the PSF parameter values: $\delta r_h^{\rm PSF}/r_h^{\rm PSF}$ (left-hand panels), $\delta e_1^{\rm PSF}$ (middle panels), and $\delta e_2^{\rm PSF}$ (right-hand panels). $\delta r_h^{\rm PSF}$ ($\delta e_i^{\rm PSF}$) is the PSF half-light radius (component of the ellipticity) used in the test set images minus the value used in the training set images (see Table 1 and Section 5 for values used in the training sets). Black (blue) points show the values obtained from noise-free (noisy) images. Black (blue) lines are linear regression fits to the noise-free (noisy) data, with dash–dotted lines showing fits to the mean of $\Delta m_1$ and $\Delta m_2$ (top left) and solid and dashed lines showing fits to $\Delta c_1$ (bottom middle) and $\Delta c_2$ (bottom right), respectively. Grey shaded regions as in Fig. 6.

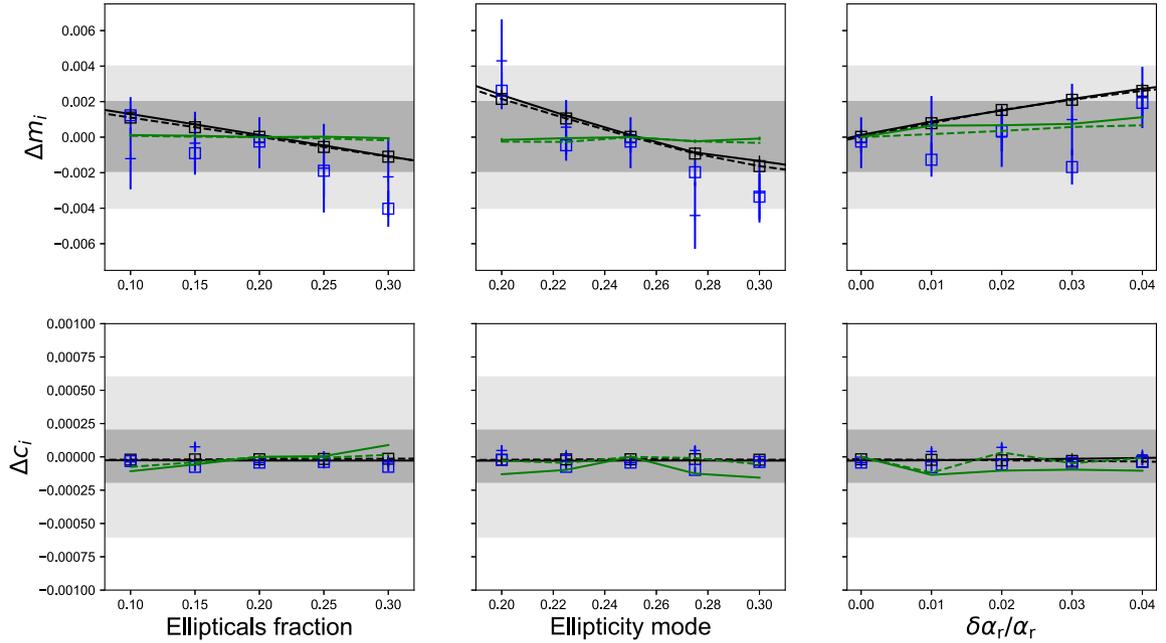

**Figure 9.** Increase in the multiplicative ($\Delta m_i$; upper panels) and additive ($\Delta c_i$; lower panels) biases with respect to the fiducial setting (see Fig. 6 and Section 8) for $i = 1$ (crosses; solid) and 2 (open squares; dashed) as a function of the galaxy distribution parameter values used in the test sets: fraction of elliptical galaxies to total number of galaxies (left-hand panels), mode of the ellipticity distribution (middle panels), and $\delta\alpha_r/\alpha_r$ (right-hand panels), where $\alpha_r$ is the slope of the size–magnitude relation (see equation (20)) used in the training sets and $\delta\alpha_r$ is the value used in the test sets minus the value used in the training sets. The galaxy distribution parameter values in the training sets are described in Section 5 and summarized in Table 1. Black (blue) points show the values obtained from noise-free (noisy) images, with lines joining the noise-free results. For comparison, green solid (dashed) lines show the biases for noise-free simulations when we used the same (offset) parameters in both the training and the test sets. Grey shaded regions as in Fig. 6.

We check that the larger biases are caused by the differences between the galaxy populations in the test and training sets, rather than being inherent to the shifted galaxy size and *S/N* distributions (shown in Fig. 4), by plotting the biases when we use the same shifted distributions in both the training and the test sets. The results, shown in Fig. 9, imply that the biases indeed result from the galaxy distribution bias, rather then from the distributions themselves. In addition, we find that the distribution biases do not reduce when we increase the training set size by a factor of five to $n_{gal} = 5 \times 10^5$, even in the noise-free case.

### 9.2 Morphology bias

Here, we look briefly at the sensitivity of the CNN shear estimates to morphology bias, in which the model is insufficient to describe observed galaxy light profiles. We begin by simulating a population of galaxies in the test sets using Sérsic indices with fixed offsets from the values used in the training sets e.g. for an offset of +0.1 (−0.1), the Sérsic indices used in the test sets are 1.1 (0.9) and 4.1 (3.9) for the disc and elliptical galaxies, respectively (see Table 1 for the values used in the training sets). Results are shown both for noise-free and noisy images in Fig. 10. We find that the multiplicative biases are significant and decrease (increase) if the Sérsic index is larger (smaller) in the test sets than in the training sets. For the noise-free simulations, there is a relatively larger increase in the magnitude of the bias when the Sérsic index is smaller (as opposed to larger) than the corresponding training set value. For the galaxy population adopted here, we find that the morphology bias is smaller in magnitude when the images are noisy. We use the correct PSF in the training sets, and thus, as expected, the additive biases are relatively unaffected.

We explore the morphology bias further by simulating galaxy intensity profiles in the test sets with a *range* of Sérsic indices. Specifically, we draw galaxies for the test sets from a uniform distribution, increasing the $n_s$ range from ±0.05 to ±0.65 around the central values $n_s = 1$ and 4. Results are shown in Fig. 10. We see that the biases measured for the noise-free images remain below the *Euclid* requirements for Sérsic indices within approximately ±0.2 of the values adopted in the training sets. Notably, the biases found for the noisy images are approximately flat for a wide range in $n_s$ values around those used in the training sets (∼±0.5); this relative insensitivity to the galaxy intensity profiles is encouraging, though we caution that the bias for a particular survey will depend on the true distribution of Sérsic profiles relative to those used to build the CNN, as well as on the actual observed galaxy morphologies, which are more complex than the single-component, elliptical isophote profiles we consider here. We discuss the issue of complex galaxy morphologies and further tests of the morphology bias that would build on this work in Section 10. Finally, we see a clear interaction between the morphology bias and the presence of noise, with the biases for noisy images inconsistent with those found for noise-free images for both an '$n_s$ offset' between the training and test sets and an '$n_s$ range' in the test sets.

## 10 DISCUSSION AND FUTURE WORK

Measuring galaxy shear with the accuracy required for next-generation surveys is a non-trivial task that has been extensively addressed by the weak lensing community, including collaborative efforts to test existing pipelines. However, few methods (e.g. Huff & Mandelbaum 2017) meet the stringent requirements on systematics that are needed to fully realize the potential of these surveys. It is crucial therefore that novel methods are developed, as well as existing methods refined, and that these are used to compare and verify shear estimates from different shape measurement pipelines. More recently, ML and, in particular, ANNs, have been applied to this task, with promising results (Gruen et al. 2010; Ribli, Dobos & Csabai 2019; Tewes et al. 2019; Zhang et al. 2023). In this work, we have explored the potential of CNNs in precision shear measurement; in particular, employing a shallow network and an MSB loss function, we have quantified the sensitivity of shear biases to the accuracy of the PSF model and, separately, the fidelity of the galaxy population, simulated in the training sets.

For the PSF model, in order to meet the shear bias requirements for *Euclid* (see Section 2.2), we find that: (i) each component of the ellipticity, $e_i^{PSF}$, must be accurate to within $10^{-3}$ and (ii) the relative absolute error in the half-light radius, $|\delta r_h^{PSF}|/r_h^{PSF}$, must be less than 0.5 per cent. Quantifying how accurately the PSF must be known, in terms of its size, ellipticity and profile shape is a primary driver to telescope design (e.g. Paulin-Henriksson et al. 2008; Cropper et al. 2010; Massey et al. 2013; Racca et al. 2016). We compare our results to the requirements on the knowledge of the PSF set out in *Euclid*'s Definition Study Report (DSR; Laureijs et al. 2011, see their table 3.5) and quoted recently in Liaudat, Starck & Kilbinger (2023).[15] Converting between ellipticity and size measured in the DSR using quadrupole moments and the definitions we use here[16] (see Sections 2.1 and 3.1), these translate to $|\delta e_i^{PSF}| < 10^{-4}$ and $|\delta r_h^{PSF}|/r_h^{PSF} < 5 \times 10^{-4}$. Thus, the requirements on the PSF model accuracy found in our simulations are considerably (by an order of magnitude) less stringent than those documented in *Euclid*'s DSR.

We caution, however, that the lower cut-off to the galaxy sizes we adopt ($r_h \geq 0.2$ arcsec; see Section 5) corresponds to a larger PSF-convolved galaxy FWHM to PSF FWHM for the disc galaxies than is quoted in the DSR.[17] Furthermore, the bounds on $\delta e_i^{PSF}$ and $\delta r_h^{PSF}/r_h^{PSF}$, we infer from our simulations are based on *Euclid*'s total error budget and thus, in reality, will need to be stricter once other sources of bias (see Section 1) are also taken into account. In addition, we have made several simplifying assumptions concerning both the PSF (see below) and the galaxies (which we discuss later in this section).

---
[14]Note that we refer to the 'steepness' of the slope given in equation (20), not the gradient of $r_h$ as a function of $m_{AB}$.

[15]RMS error on each ellipticity component, $|\delta \epsilon_i^{PSF}| < 2 \times 10^{-4}$ and relative RMS error on the size, $|\delta R_{PSF}^2|/R_{PSF}^2 < 10^{-3}$, where $\epsilon^{PSF}$ and $R_{PSF}^2$ are, respectively, the ellipticity and size measured using quadrupole moments (see, for e.g. Paulin-Henriksson et al. 2008).
[16]$\delta \epsilon_i^{PSF} \approx 2\delta e_i^{PSF}$ and $\delta R_{PSF}^2/R_{PSF}^2 \approx 2\delta r_h^{PSF}/r_h^{PSF}$.
[17]In the DSR, the sample is quoted as being restricted to galaxies with FWHM 1.25 times larger than that of the PSF.

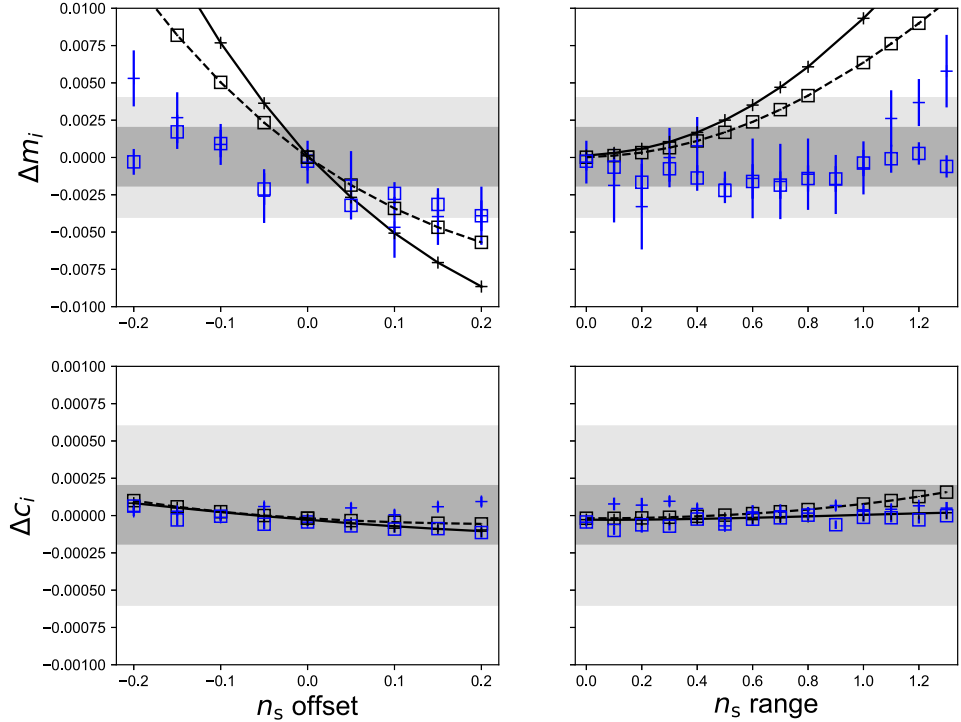

**Figure 10.** Increase in the multiplicative ($\Delta m_i$; upper panels) and additive ($\Delta c_i$; lower panels) biases with respect to the fiducial setting (see Fig. 6 and Section 8) for $i = 1$ (crosses; solid) and 2 (open squares; dashed) as a function of (i) the Sérsic index used in the test sets minus the value used in the training sets (left; $n_s$ offset) and (ii) the range of Sérsic indices used to represent disc (elliptical) galaxies in the test sets (right; $n_s$ range), with Sérsic indices drawn from a uniform distribution around $n_s = 1$ (4). See text in Section 9.2 for further details. Black (blue) points show the values obtained from noise-free (noisy) images, with lines joining the noise-free results. Grey shaded regions as in Fig. 6.

In this paper, we consider a non-varying PSF and quantify the requirements for the accuracy of the PSF model parameter values adopted in the training sets. However, in reality, the PSF varies spatially across the FoV and over time and also depends on the galaxy SED. A practicable CNN shear measurement pipeline will need to address this issue, for example, by training a suite of CNNs, with each individual network built using a different PSF.

Spatio-temporal effects on the PSF shape are typically captured using observations of stars in the field and interpolating to the positions of the galaxies. Refining current methods for reconstructing the PSF from stars will be important to ensure that requirements are met (e.g. Schmitz et al. 2020). Forward-modelling approaches using ray-tracing through the telescope optics have also been adopted (see Mandelbaum 2018, and references therein). In terms of spectral dependence, in addition to accurate modelling of the instrumental PSF as a function of wavelength, individual galaxy SEDs must be estimated sufficiently well. Eriksen & Hoekstra (2018) explore this issue and conclude that it is possible to achieve *Euclid*'s accuracy requirements on the PSF size using photometric data. We note that the observed link between galaxy colour and morphology (e.g. Masters et al. 2019; Uzeirbegovic, Martin & Kaviraj 2022) could be utilised in the context of PSF size estimation for weak lensing shear estimation.

We also explore the sensitivity of shear estimates to the fidelity of the galaxy population used to build the CNN model, considering separately the impacts from galaxy distribution bias and morphology bias. We find that the multiplicative biases can be significant, depending on how well the training sets represent observed galaxies. In future work, it will be important to simulate more realistic galaxy morphologies, for example, including a wider range of Sérsic profiles, representing different galaxy types, as well as taking into account the complicating effects of non-elliptical isophotes in bulge plus disc galaxies, and including asymmetrical features and substructures. In particular, the widely-adopted, publicly available software package GALSIM[18] (Rowe et al. 2015) can be used to simulate galaxy images from real *Hubble Space Telescope* (*HST*) data, as well as from simple parametric models. Furthermore, we include a galaxy size–magnitude relation, but it is known that correlations exist between several galaxy properties, including, for example, a dependence of half-light radius on magnitude and Sérsic index (Euclid Collaboration: Martinet et al. 2019) and an evolution of galaxy type with redshift. If ignored in the training sets, these correlations may have a significant impact on the biases. Simulating realistic galaxy populations and quantifying the potential impact from galaxy distribution and morphology bias will be crucial to shear measurement pipelines using ANNs.

ML has been applied to galaxy classification by morphological type since the early 1990s (Storrie-Lombardi et al. 1992), using both a range of classic (e.g. Vavilova et al. 2021) and deep learning models, including CNNs (Cheng et al. 2020, and references therein). Recently, Li et al. (2022) have developed a CNN to output the Sérsic profile parameters of galaxies from seeing-limited ground-based observations. Furthermore, deep learning has been used to generate realistic galaxy images using deep generative models (Euclid Collaboration: Bretonnière et al. 2022). Both of these ML applications may play a role in the development of an accurate and precise shear measurement pipeline using ANNs. In particular, not only to simulate images for training and testing the model, but also to enable galaxy classification prior to shear estimation, allowing a separate CNN to be trained for each galaxy type.

---

[18] https://github.com/GalSim-developers/GalSim

## ACKNOWLEDGEMENTS

We thank the anonymous referee for comments that led to a significantly improved paper. We are grateful to Stuart Newman for assistance with the High Performance Computer at the University of Essex.

## DATA AVAILABILITY

The data underlying this article will be shared on reasonable request to the corresponding author.

This paper has been typeset from a TeX/LaTeX file prepared by the author.